\documentclass[aps,prb,twocolumn,showpacs,superscriptaddress]{revtex4}
\usepackage{graphicx}
\usepackage{amsfonts,amssymb,amsmath}
\usepackage{bm}
\begin{document}

\title{An {\it ab-initio} investigation of magnetism in two-dimensional
uranium systems}

\date{\today}
 
\author{Matej Komelj}
\email{matej.komelj@ijs.si}
\affiliation{Jo\v zef Stefan Institute, Jamova 39, SI-1000 Ljubljana,
  Slovenia}
\author{Nata\v sa Stoji\' c}
\affiliation{Abdus Salam International Centre for Theoretical Physics,
Strada Costiera 11, I-34014 Trieste, Italy}
\begin{abstract}
The orbital and spin magnetic moments, and the 
X-ray-magnetic circular-dichroism (XMCD) spectra at the $M_{4,5}$ edges of the 
U atoms in a UAs/Co multilayer and in an $\alpha$-U film
are calculated within the framework of the density-functional theory in  
combination with the local-spin-density approximation (LSDA), the 
generalized-gradient approximation (GGA) and the LDA+$U$ method.  
The antiparallel arrangement between the U and Co spin magnetic moments at the
interface results in the vanishing  of ferromagnetism for the case of very thin
Co layers. The U moments decay rapidly with the distance from the
film surface. The magnitude of the magnetic-dipole term $\langle T_z\rangle$, 
which appears in the spin XMCD sum rule, is small. 
The different exchange-correlation treatments do not yield qualitatively 
different results.
\end{abstract}
\maketitle
Both the subject of actinides and the issue of magnetism in two dimensional
systems, like multilayers or films, represent interesting topics in 
contemporary solid-state physics. The joint problems from the two fields
are therefore even more challenging. Actinides and actinide compounds
exhibit very different magnetic behaviors, such as Pauli paramagnetism,
localized and itinerant magnetism and heavy 
fermions.\cite{Freeman:1984,Lander:1991} It is expected that the properties
would change when going from bulk to two-dimensional materials. 
For example, Plasket {\it et al.}\cite{Plaskett:1992} produced  UAs/Co 
multilayers  in an attempt to increase the Curie temperature of the amorphous 
UAs/Co alloy due to the induced polarization of U atoms through the exchange
coupling to the Co layers. The first results, which came from
magneto-optical measurements, were promising, but they were not reproduced 
with polarized-neutron-reflectivity experiments\cite{Mannix:1997} because 
the main contribution to the measured signal arises from the Co atoms. 
As an alternative approach, Kernavanois {\it et al.}\cite{Kernavanois:2004}
performed a measurement of the X-ray-magnetic-circular dichroism (XMCD), which  
is element specific\cite{Schutz:1987}. The same technique may be suitable for
detecting the finite magnetic moment at the surface of uranium that was
predicted theoretically\cite{Stojic:2003}.

The XMCD spectroscopy is based on the difference in the absorption coefficients
$\mu^+(\epsilon)$, $\mu^-(\epsilon)$ and $\mu^0(\epsilon)$ for circularly-right,
circularly-left and linearly polarized X-rays.  The $5f$ contributions  to the
magnetic orbital $m_l=-\mu_B\langle l_z\rangle$ and spin $m_s=-\mu_B\langle
\sigma_z\rangle$ moments are related to the absorption coefficients 
at the $M_{4,5}$ edges,  via the  sum rules\cite{Thole:1992,Carra:1993}:
\begin{eqnarray}
\label{orbsumrule} \langle l_{z}\rangle & = &
\frac{3I_{\text{m}}N_{\text{h}}}{I_{\text{t}}} \quad , \\
\label{spinsumrule} \langle \sigma_{z}\rangle & = &
\frac{3I_{\text{s}}N_{\text{h}}}{I_{\text{t}}} - 3\langle T_{z}\rangle
\quad , \\ 
I_{\text{m}} & = & \int \left[
(\mu_{\text{c}})_{M_5} + (\mu_{\text{c}})_{M_4} \right] d\epsilon \quad , \\
I_{\text{s}} & = & \int \left[
(\mu_{\text{c}})_{M_5} - {3\over 2}(\mu_{\text{c}})_{M_4} \right] d\epsilon \quad ,\\ 
I_{\text{t}} & = & \int \left[
(\mu_{\text{t}})_{M_5} + (\mu_{\text{t}})_{M_4} \right] d\epsilon \quad ,
\end{eqnarray}
with the XMCD signal $\mu_{\text{c}}=\mu^+-\mu^-$ and the total 
absorption coefficient $\mu_{\text{t}}= \mu^++\mu^-+\mu^0$. The symbols 
$(\mu_{\text{c,t}})_{M_{4,5}}$ in Eqs. (3)-(5) denote that the integration is
to be performed over the energy integral corresponding to the $M_{4,5}$ edges.
There are some problems connected with the application
of the sum rules on experimental data. The exact number of holes $N_\text{h}$
in the $f$ band is not directly measurable, nor is the 
expectation value  
$\langle T_z\rangle$ of the magnetic
dipolar operator:
\begin{equation}
\hat{T}_{z} = \frac{1}{2} [ \bm{\sigma} -
3\hat{\mathbf{r}}(\hat{\mathbf{r}}\cdot\bm{\sigma}) ]_{z} \quad ,
\end{equation}
where $\bm{\sigma}$ denotes the vector of the Pauli matrices. 
The integrations in Eqs. (3)-(5) have to be performed over the energy interval
which corresponds to the electron $3d\rightarrow 5f$ transition. 
It is important to note that the limits of this energy interval are not uniquely 
defined in the experiment and that it is assumed
that the  electron $3d\rightarrow 5f$ transition is the only relevant 
contribution to the absorption signal in this energy range.
Above all, the sum rules are derived for the atomic case. 
In spite of all these limitations, the XMCD analysis,
particularly at the $L_{2,3}$ edges, has
been proven to be a very successful tool for investigations of the magnetism 
in transition-metal systems of various dimensionalities. 
However, there is less research about the validity of the sum rules for other materials,
for example, actinides, especially in systems other than bulk.  Hence, the 
subject of the present paper is a comparison 
of the magnetic moments calculated directly from the electronic structure with 
the moments obtained from the theoretical XMCD spectra by using Eqs. 
(\ref{orbsumrule},\ref{spinsumrule}). The theoretical approach has several 
advantages over an experiment since some of the above-mentioned uncertainties,
particularly ones related to the number of holes, the $\langle T_z\rangle$ term  
and the integration limits, can be reduced to a minimum. This enables us to 
investigate the validity of the sum rules
(\ref{orbsumrule},\ref{spinsumrule}) without the influence
of other effects present in the experiment, which are hard to control. The number of 
holes $N_\text{h}$ is well defined. The integrations in Eqs. (3)-(5) are 
carried out
over the interval between the Fermi energy $E_\text{F}$ and the upper limit
$E_\text{C}$,
which is related to the $N_\text{h}$ and to the $f$-resolved density of states
$n_f(\epsilon)$ as $N_\text{h}=\int_{E_\text{F}}^{E_\text{C}} n_f(\epsilon)\>
d\epsilon$. The $\langle T_z\rangle$ term is simply calculated as the 
expectation value of (6). 

The calculations  were performed within the framework of the density-functional
theory by applying the Wien97 code \cite{Blaha:1990}, which adopts the
full-potential linearized-augmented-plane-wave (FLAPW) method\cite{Wimmer:1981}.
The Brillouin-zone (BZ) integrations were carried out with the 
modified tetrahedron method\cite{Blochl:1994} by using 600 and 550 $k$ points
for the UAs/Co multilayer and the U film, respectively. The plane-wave
cut-off parameter was set to $11.4\>{\rm Ry}$ for both systems.
In order to investigate the influence of correlation effects, the 
exchange-correlation contribution to the effective potential was calculated
by using different schemes, namely
the local-spin-density approximation (LSDA)\cite{Perdew:1992-1}, 
the generalized-gradient approximation (GGA)\cite{Perdew:1992-2} and
the LDA+$U$ method\cite{Anisimov:1993,Liechtenstein:1995}.
The parameters $U$ and $J$, which appear in the LDA+$U$ scheme, were set to
$U=2\>{\rm eV}$ and $J=0.5\>{\rm eV}$\cite{Yaresko:2003}.
The contribution of the spin-orbit coupling (SOC) was calculated in the
second-variational scheme\cite{Singh:1994,Novak?}, since it was demonstrated
that a more rigorous 
treatment,  for example, the first-variational scheme\cite{Komelj:2004}, did not
yield substantially different results for uranium, although the magnitude
of the $5f$ uranium SOC interaction is comparable to the corresponding bandwidth.
The  absorption coefficients  $\mu^+(\epsilon)$, $\mu^-(\epsilon)$ and
$\mu^0(\epsilon)$ as a function of the photon energy $\epsilon$
were calculated using Fermi's golden rule in a nonrelativistic dipole 
approximation that is based on the evaluation of the matrix element
for the operator $\hat{\mathbf p}\cdot \mathbf{e}$, with
$\mathbf{e}$ denoting the polarization vector of the light\cite{Wu:1994,
Kunes:2001-1,Kunes:2001-2}. However, note that the current 
implementation of the density functional theory cannot simulate the 
excited states which would result from a hole in the $3d$ state.

The UAs/Co multilayer was modelled with a supercell, based on a 
face-centered-cubic (fcc) stacking sequence along the (001) direction as
presented in Fig. 1. The experimental lattice parameter  
$a=0.355\>{\rm nm}$ for fcc Co was used. It turns out that the calculations 
performed on supercells with fewer Co layers do not result in a magnetic 
solution, in agreement with experiments at $300\>{\rm K}$ (Ref. 
\onlinecite{Plaskett:1992}),
where no magnetic ordering
was observed for the multilayers with the Co layer thinner than $1\>{\rm nm}$.
However, the present theoretical results do not prove the thesis that the 
absence of magnetism is due to the interdiffusion of U and As atoms into the 
Co layer.  
\begin{figure}
\includegraphics[width=0.45\textwidth]{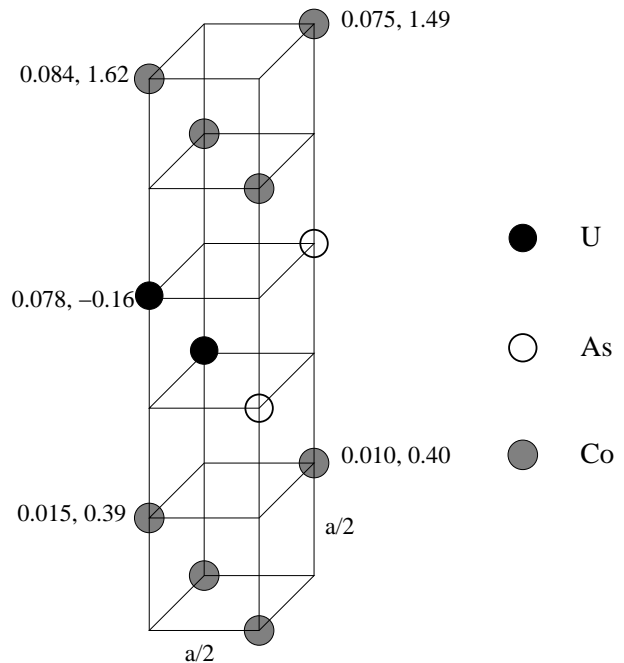}
\caption{A part of the As/Co-multilayer supercell, which is based on the fcc
Co unit cell with the lattice parameter $a$. The numbers
$\langle l_z\rangle, \langle\sigma_z\rangle$ 
represent
the spin and orbital moments at particular U and Co sites obtained by applying the 
LDA+$U$ method for the $5f$ U states.}
\end{figure}
Furthermore, in the case of larger supercells, with more UAs layers, a finite
magnetic moment at the U sites was found just at the interface, while in 
experiments\cite{Kernavanois:2004} the magnitude of the moment
increased with an increasing UAs thickness. 
The calculated spin $\langle\sigma_z\rangle$ and orbital $\langle l_z\rangle$ 
magnetic moments, obtained with
the LDA+$U$ method, are presented in Fig. 1. 
As demonstrated in Table 1
for the case of the U $5f$ moments,
the LSDA and GGA do not give qualitatively different results,
although the Coulomb repulsion $U$, in terms of the LDA+$U$ method, enhances the 
orbital moments by  about a factor of 2 (3) relative to the LSDA (GGA) values.
In contrast to the low-temperature experiments\cite{Kernavanois:2004} on
the multilayers with amorphous UAs bilayers, carried out in the presence of high
magnetic fields, the magnitude of the orbital moments are smaller than
the magnitude of the spin moments. The magnitude of the total magnetic moment
$\left|\langle l_z\rangle+\langle\sigma_z\rangle\right|\approx 0.1$ is 
much smaller than the corresponding experimental values\cite{Kernavanois:2004}
between $0.4$ and $1.1$.  The discrepancy might be partly due to the difference
in the ratio of atomic concentrations, [U]/[As], between the calculation ([U]/[As]=1) and the
experiment\cite{Kernavanois:2004} ([U]/[As]=1.5) since the results of the SQUID 
measurements\cite{Fumagalli:1992} exhibited a pronounced composition dependence.
At the interface there is an antiparallel alignment between the calculated 
U and Co spin moments. The
Co moments in the interface layer are drastically reduced relative to the 
moments in other layers or in the bulk, which  is consistent with the vanishing of 
magnetism in the case of thin Co layers, as mentioned earlier. \par
A comparison between the U $5f$ moments calculated directly from the 
electronic structure and the moments obtained from the theoretical XMCD  spectra
is given in Table 1.
\begin{table*}
\begin{ruledtabular}
\begin{tabular}{lccccc}
&$\langle l_z\rangle$&$\langle l_z\rangle$ from Eq. (\ref{orbsumrule})&
$\langle \sigma_z\rangle$&$\langle \sigma_z\rangle$ from Eq.  
(\ref{spinsumrule})&
$\langle \sigma_z\rangle$ from Eq.  (\ref{spinsumrule}), $\langle T_z\rangle=0$
\\
\hline
LSDA   &0.033&0.034&-0.138&-0.105&-0.109\\
GGA    &0.025&0.027&-0.169&-0.121&-0.104\\
LDA+$U$&0.078&0.068&-0.162&-0.128&-0.138
\end{tabular}
\end{ruledtabular}
\caption{A comparison between the $5f$ contribution to the orbital 
$\langle l_z\rangle$ and the
spin $\langle\sigma_z\rangle$ magnetic moments at the U sites in the UAs/Co
multilayer, calculated
directly from the electronic structure, and obtained from the theoretical
XMCD spectra by using the sum rules (\ref{orbsumrule},\ref{spinsumrule}).}
\end{table*}
The orbital sum rule (\ref{orbsumrule}) holds well for all three types of the 
exchange-correlation potential. The validity of the spin sum rule 
(\ref{spinsumrule}) is worse, since the resulting moments underestimate
the directly calculated values by $\sim 20\%$ (LDA+$U$) to $\sim 30\%$ (GGA).
The influence of the magnetic dipole $\langle T_z\rangle$ term is below
the deviation of the sum rule $\langle\sigma_z\rangle$ from the
directly calculated value although, surprisingly, the deviation is reduced
when the $\langle T_z\rangle$ is omitted from Eq. (\ref{spinsumrule}) in the
case of the LSDA and LDA+$U$.

The supercell for modelling the $\alpha$-U (001) film was based on the orthorhombic 
structure and it contained 6 uranium and 4 vacuum layers (for details, see,
for example, Ref. \onlinecite{Stojic:2003}). As found 
previously\cite{Stojic:2003}, the density-functional theory predicts a magnetic
ordering on the U surface. 
\begin{figure}
\begin{center}
\vspace{36pt}
\includegraphics[width=0.45\textwidth]{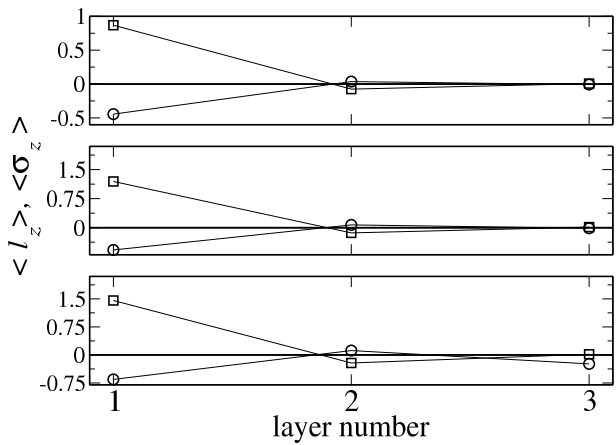}
\caption{The $5f$ orbital $\langle l_z\rangle$ (circles) and the spin 
$\langle\sigma_z\rangle$ (squares) magnetic moments as a function of the 
layer number for U film, calculated with the LSDA (upper graph), GGA 
(middle graph) and LDA+$U$ method (lower graph).}
\end{center}
\end{figure}
Fig.  2 demonstrates that the calculated magnetic moments rapidly decrease
with an increasing distance from the surface. However, the sign of the moments
alternates, so that the moments in subsequent layers are aligned antiparallel
to each other. In the surface and in the next layer, where the moments
are possibly measurable, the magnitude of the spin moment 
$\langle\sigma_z\rangle$ is about two times
larger than the magnitude of the orbital moment $\langle l_z\rangle$.  The
enhancement of the magnitude of $\langle l_z\rangle$ due to the Coulomb
repulsion is smaller than in the case of the UAs/Co multilayer, whereas the 
LDA+$U$ spin magnetic moment is almost two times larger than the corresponding
LSDA value.
The calculated XMCD signal is pronounced just for the surface layer, as  
shown in Fig. 3.
\begin{figure}
\begin{center}
\vspace{36pt}
\includegraphics[width=0.45\textwidth]{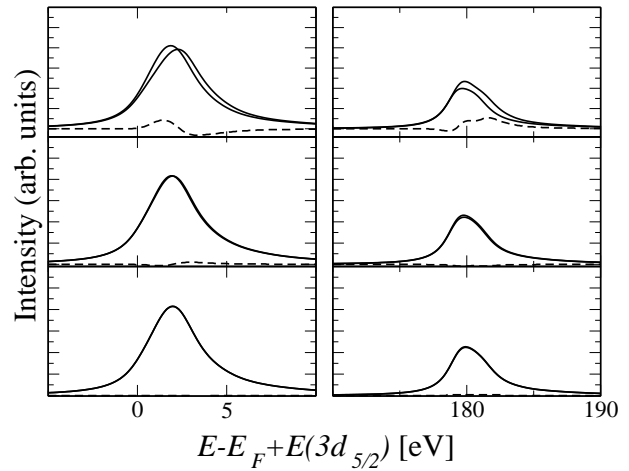}
\caption{The calculated absorption spectra for two polarizations (solid lines) and the XMCD
signal (dashed lines) for the surface (upper graph), the second (middle graph) 
and the third (lower graph) layer of U film obtained by using the LDA+$U$ 
method. The $M_5$ edge is shown on the left and the $M_4$ edge on the right panel.}
\end{center}
\end{figure}
Hence, only the surface magnetic moments are given in Table 2. 
\begin{table*}
\begin{ruledtabular}
\begin{tabular}{lccccc}
&$\langle l_z\rangle$&$\langle l_z\rangle$ from Eq. (\ref{orbsumrule})&
$\langle \sigma_z\rangle$&$\langle \sigma_z\rangle$ from Eq.  
(\ref{spinsumrule})&
$\langle \sigma_z\rangle$ from Eq.  (\ref{spinsumrule}), $\langle T_z\rangle=0$
\\
\hline
LSDA   &-0.444&-0.364&0.869&0.722&0.714\\
GGA    &-0.576&-0.479&1.193&0.981&0.956\\
LDA+$U$&-0.655&-0.546&1.456&1.194&1.162
\end{tabular}
\end{ruledtabular}
\caption{A comparison between the $5f$ contribution to the orbital 
$\langle l_z\rangle$ and the
spin $\langle\sigma_z\rangle$ U magnetic moments at the surface of the
U film calculated
directly from the electronic structure, and obtained from the theoretical
XMCD spectra by using the sum rules (\ref{orbsumrule},\ref{spinsumrule}).}
\end{table*}
Both the orbital and the spin sum rule underestimate the directly
calculated magnetic moments. The deviation is below $20\%$.
The influence of the $\langle T_z\rangle$ term is even more subtle than in 
the case of the UAs/Co multilayer, although the validity of the spin
sum rule is worse if this term is set to zero.

In conclusion, a theoretical test based on an {\it ab-initio} calculation 
proved the
XMCD method to be a suitable tool for investigating the magnetism in 
two-dimensional U systems, that arises from the ordered $5f$ magnetic moments.
However, the results of the calculations on the UAs/Co multilayer suggest the 
ferromagnetism confined at the interface, and hence to some extent disagree 
with the experimental data and findings. 
If ferromagnetism really exists on the surface of uranium, its presence should be
readily proven by means of the XMCD measurements.\par
We are grateful to N. Binggeli for the reading of manuscript and her 
suggestions. 
\bibliography{uranium.bib}
\end{document}